\documentclass{article}
\usepackage{graphicx}
\usepackage[
backend=biber,
style=nature,
doi=false,isbn=false,url=false,eprint=false
]{biblatex}
\usepackage{geometry}
\usepackage{pdflscape}
\usepackage{graphicx}
\usepackage{array} 
\usepackage{varwidth}

\title{Equity considerations in COVID-19 vaccine allocation modelling: a literature review }
\author{Eva Rumpler, Marc Lipsitch}

\begin{document}

\maketitle

\section{Abstract}

We conducted a literature review of COVID-19 vaccine allocation modelling papers, specifically looking for publications that considered equity. We found that most models did not take equity into account, with the vast majority of publications presenting aggregated results and no results by any subgroup (e.g. age, race, geography, etc). We then give examples of how modelling can be useful to answer equity questions, and highlight some of the findings from the publications that did. Lastly, we describe seven considerations that seem important to consider when including equity in future vaccine allocation models.

\section{Introduction}

As COVID-19 vaccines became available in 2021, countries and regions around the globe had to define criteria for prioritizing access to vaccines. 
For example, WHO’s Strategic Advisory Group of Experts on Immunization (SAGE) released a roadmap in October 2020 for prioritizing uses of COVID-19 vaccines in the context of limited supply. In situations of community transmission, they recommended vaccines be first given to healthcare workers and older individuals, followed by groups (defined by health states or sociodemographic status) at significantly higher risk of severe disease and deaths \cite{world_health_organization_who_2021}. 
Similarly in the US, the Advisory Committee on Immunization Practices recommended vaccines be allocated first to health care personnel and residents of long-term care facilities, followed by older individuals, people with high-risk medical conditions, and essential workers not already included \cite{dooling_advisory_2020}. \\

\noindent Some have argued that vaccine allocation should provide equitable protection of individuals and should aim to reduce (or at least not further increase) inequities in health burden. 
For example, WHO's SAGE developed a values framework for the allocation and prioritization of COVID-19 vaccines, defining six core principles for equitable protection. In addition to human well-being, equal respect, reciprocity and legitimacy, two of the principles were global equity (i.e. vaccine access to all counties, particularly low- and middle-income countries) and national equity (i.e.  preferential allocation for vulnerable groups and groups experiencing higher burden) \cite{world_health_organization_who_2020}. 
Similarly, the US National Academies of Sciences, Engineering, and Medicine, in its 2020 Framework for equitable allocation of COVID-19 vaccine defined three ethical principles: maximum benefit, equal concern, and mitigation of health inequities, defined as the obligation to explicitly address the higher burden of COVID-19 experienced by some populations \cite{gayle_framework_2020}. \\

\noindent Modelling has provided useful contributions to identifying the best performing allocation strategies for COVID-19 vaccines and has helped inform policy making in the face of uncertainty about the vaccines and pathogens characteristics. Many publications have modelled the performance of different allocation strategies, often comparing transmission-minimizing approaches targeting younger adults with mortality-minimizing approaches targeting older age groups when vaccine stock is limited \cite{bubar_model-informed_2021, hogan_report_2020, matrajt_vaccine_2020, moore_modelling_2021}. \\

\noindent Despite huge inequities in disease burden distribution and large modelling literature on allocation optimization, however, few publications modelled equity considerations. Here we present a literature review conducted to identify publications that did look at equity and COVID-19 allocation modelling. We highlight some of their main findings and summarize advice to modellers wanting to include equity in future models. \\

\section{Methods} 

We conducted a literature review and expert-opinion elicitation in search of papers that looked at COVID-19 vaccine allocation, presenting outcomes stratified by any characteristic (age, race, ethnicity, occupation, geography, etc). \\

\noindent We performed a literature review using PubMed, including studies published between 2019 and May 2024. We included all peer-reviewed publications and pre-prints that were relevant to our search terms for:  
\begin{itemize}
    \item COVID-19 : "COVID-19 Vaccines"[MeSH] OR "SARS-CoV*"[ti] OR COVID*[ti] OR coronavir*[ti] OR NCOV[ti]
    \item Vaccine: vaccine*[ti] OR vaccination[ti]
    \item Modelling: "Computer Simulation"[Mesh] OR "Epidemiological Models"[Mesh]  OR "Models, Statistical"[Mesh] OR model*[ti] OR scenario[ti] OR strategies[ti] OR strategy[ti] OR prioritization[ti]
    \item Disparity or subpopulation: "Demography"[Majr] OR "Population Groups, US"[Mesh] OR "Age Factors"[Mesh] OR "Race Factors"[Mesh] OR "Aged"[Mesh] OR demographic*[ti] OR equit*[tiab] OR inequit*[tiab] OR "fair distribution"[tiab:~3] OR "fair allocation"[tiab:~3] OR race[ti] OR racial*[ti] OR ethnic*[ti] OR age[ti] OR aged[ti] OR elder*[ti] OR "risk factor*"[ti]
\end{itemize}

\noindent From this search, we identified 251 papers to be reviewed. After review, 6 of them corresponded to our criterion for inclusion. \\

\noindent To look for additional relevant studies, we carried out a manual search of all studies cited or citing the papers we had already identified and found an additional 6 publications. \\ 

\noindent Finally, having found only a limited number of research papers that meet our criteria in the literature review, we conducted an expert-opinion elicitation to identify research papers that we may have missed. We contacted four experts in public health and the study of disparities, asking if they knew of publications matching our criteria. All provided publications we had identified already. The experts were (in alphabetical order) Dr. Mary Bassett (MD, MPH), Harvard University, Dr. Mathew Kiang (MPH, ScD), Stanford University, Dr. Andrew C. Stokes (MA, PhD), Boston University, and Dr. Elizabeth Wrigley-Field (MA, PhD), University of Minnesota. \\

\noindent Our literature review has limitations. Firstly, although we searched for the most publications possible by doing both a literature review and expert-opinion elicitation, there may be publications that we missed. It can be especially difficult to identify publications as not all explicitly mention equity, and some use diverse vocabulary to describe similar analyses. 
Secondly, we focused our literature review on prioritization of COVID-19 initial vaccination. We did not include publications regarding allocation of boosters as these took place under less vaccine quantity constraints.

\section{Literature review findings}

\noindent Of the numerous publications that modelled COVID-19 vaccine allocation, most did not explicitly consider equity. Despite an extended literature search directed at literature considering equity between population subgroups and manual review of 251 publications, only 6 (2.4\%) \cite{aiona_disparate_2022, chapman_risk_2022, kadelka_ethnic_2022, kipnis_evaluation_2021, rosenstrom_can_2022, rumpler_fairness_2023} were COVID-19 vaccine allocation modelling studies presenting results by some sub-group. From manual search among studies cited or citing those studies, we identified an additional 6 publications \cite{chen_strategic_2022, hoover_californias_2024, stafford_retrospective_2023, wrigley-field_geographically_2021, munguia-lopez_fair_2021, khodaee_humanitarian_2022}. Expert-opinion elicitation of four experts did not provide any additional publications. Of these, 11 explicitly consider equity in their model, and one \cite{chapman_risk_2022} strives to identify the optimal allocation without explicitly mentioning equity. Characteristics and key findings from these 12 publications can be found in Table \ref{tab:litreview} below. \\

\noindent Contrary to the hypothesis of an equity/efficiency trade-off, there often is no such trade-off found, or it is weaker than one might expect, as it can be possible to minimize total deaths or infections while simultaneously reducing disparities between groups in the risks of these outcomes. Multiple publications have found that the most efficient policy overall is either the one most beneficial to disadvantaged groups, or one that prioritizes disadvantaged groups \cite{chen_strategic_2022, rumpler_fairness_2023, wrigley-field_geographically_2021, kadelka_ethnic_2022, rosenstrom_can_2022, stafford_retrospective_2023}. 
Notably, Chen et al. (2022) show that prioritizing the most disadvantaged based on any criteria (age, race/ethnicity, occupation, or income) increases both equity (which they define as mitigation of mortality disparities in disadvantaged demographic group) and social utility (which they define as minimizing mortality in the entire population) \cite{chen_strategic_2022}. 
Rumpler et al. (2023) show that a vaccine allocation policy targeting older individuals saves most lives overall and among every racial/ethnic group despite leading to less vaccines allocated to non-White people than an allocation policy targeting non-healthcare front-line workers \cite{rumpler_fairness_2023}. 
Wrigley-Field et al. (2021) show that targeting racial/ethnic groups (or geography) makes vaccine available to people with higher mortality rates, and also increases the proportion of BIPOC eligible \cite{wrigley-field_geographically_2021}.
Kadelka et al. (2022) show that taking race/ethnicity into account when prioritizing vaccines may reduce the total number of deaths, especially when also assuming that proportionately more people of color occupy high-contact jobs \cite{kadelka_ethnic_2022}.
Rosenstrom et al. (2022) show that a strategy to increase vaccination uptake in non-White individuals reduces deaths overall without significantly increasing the infection rate \cite{rosenstrom_can_2022}. 
Stafford et al. (2023) find that to minimize disease burden (death), the age cutoff for vaccine eligibility should be younger for people of color than for White people, suggesting that universal age thresholds aren't the most efficient \cite{stafford_retrospective_2023}.
\noindent While there are numerous instances of policies that improve equity and efficiency simultaneously compared to alternatives, we note that definitions of equity can vary and that maximizing one aspect may not maximize all. We discuss this further below. \\

\noindent 
Researchers face trade-offs in formulating research questions and performing the analyses to answer these questions\\ 
\noindent
A first trade-off is between considering allocation policies that are complex and may be optimal, versus considering simpler policies that may be more realistic. Some publications looked at allocation rules that consider only one or two characteristics (e.g. race, age, occupation, etc) \cite{rumpler_fairness_2023, aiona_disparate_2022, wrigley-field_geographically_2021, kipnis_evaluation_2021, rosenstrom_can_2022, stafford_retrospective_2023, munguia-lopez_fair_2021, khodaee_humanitarian_2022, hoover_californias_2024},
whereas other studies looked for the one optimal policy over multiple stratifying variables. These usually performs better but may be more difficult to implement due to a high number of variables used to define allocation priority, and may not be generalizable to other populations with different joint distributions of predictors and outcomes \cite{chapman_risk_2022, kadelka_ethnic_2022, chen_strategic_2022}. 
While considering finer allocation rules that can vary over multiple variables can lead to more efficient allocations, as shown by Kadelka at al. (20222), increasing the number of levels within each parameters can too, lead to better performance \cite{kadelka_ethnic_2022}. \\ 
\noindent
A second trade-off may exist between political acceptability and equity. Indeed while some publications consider vaccine allocation by race/ethnicity categories directly \cite{kadelka_ethnic_2022, rosenstrom_can_2022, stafford_retrospective_2023, chen_strategic_2022}, others like Wrigley-Field et al. (2021) propose using geography-based vaccine allocation at the census-tract level as an alternative to race/ethnicity strategies \cite{wrigley-field_geographically_2021}. Multiple other papers \cite{chapman_risk_2022, munguia-lopez_fair_2021, khodaee_humanitarian_2022} consider geographical units at various levels (county, states, country, etc) as targets for vaccination. \\

\noindent
Overall, comparing results from these 12 publications is made difficult by the fact that they use different terms and vocabulary, sometimes without clear definitions. This may in part be due to the fact that they come from different literatures, for example the two publications from decision science journals \cite{munguia-lopez_fair_2021, khodaee_humanitarian_2022} contrast with some of the infectious disease modelling ones \cite{kadelka_ethnic_2022, rosenstrom_can_2022, stafford_retrospective_2023, chen_strategic_2022}. \\

\section{Discussion}

\noindent The publications \cite{rumpler_fairness_2023, wrigley-field_geographically_2021, chapman_risk_2022, aiona_disparate_2022, kadelka_ethnic_2022, kipnis_evaluation_2021, rosenstrom_can_2022, stafford_retrospective_2023, munguia-lopez_fair_2021, khodaee_humanitarian_2022, chen_strategic_2022, hoover_californias_2024} we have identified provide some insights to take into account when including equity in vaccine allocation modelling: 

\begin{itemize}

    \item Most modelling studies could show some results by subgroup, even if the focus of the paper is not equity considerations. Indeed, most publications could easily show how the outcomes are distributed by subgroups, instead of only showing total outcome in the population. This is especially true for COVID-19 models that often have to include age categories due to the exponential increase in mortality rate: results by age groups are already being generated and could easily be presented. 
    As an example of how this can be done and how it enhances the understanding of results, Chapman et al. (2022) show the distribution of vaccinations, as well as infections, death and DALYs averted in each age group when the oldest age groups ($\ge$ 60 year olds) is targeted for vaccination (after initial allocation to healthcare workers and long-term care facility residents) \cite{chapman_risk_2022}. % Table 2

    \item When modelling equity considerations in vaccine allocation, one should carefully choose the outcome to be maximized. While most publications we found aimed at limiting some form of health burden (such as cases, deaths, DALYs or years of life lost), some described the performance of each strategy in terms of doses allocated to disadvantaged groups. Kipnis et al. (2021) for example present total health outcome reduction of six different allocation strategies, but presents the results by subgroup (race/ethnicity and high neighborhood deprivation index) in terms of doses allocated \cite{kipnis_evaluation_2021}. An example of an alternative can be found in Hoover et al. (2024), where the impact of California's COVID-19 vaccine equity policy is shown on both vaccine administered and COVID-19 cases \cite{hoover_californias_2024}. In some situations, maximizing the number of vaccine doses given to a group does not necessarily maximize the number of lives saved overall or in that group \cite{rumpler_fairness_2023}. The possibility of such an outcome underscores the importance of reporting both allocations and outcomes by group and giving a rationale for what is being maximized.

    \item Another aspect to take into account in setting up a modelling study for vaccine allocation is choosing which subgroup to be maximized over. In a situation where there are two predictors of the outcome that are negatively correlated to one another, maximizing the outcome over the weaker of the two predictors can lead to making the wrong allocation in terms of total outcome. This is for example likely to be the case when comparing allocations by age groups and by race/ethnicity in the US. Indeed, as illustrated in the case study \cite{rumpler_fairness_2023}, those two variables are both predictors of the outcome (in the US, older individuals, as well as Hispanic, Black, Native and Pacific Islander have higher COVID-19 death rates), and negatively correlated to one another (in the US, older individuals are disproportionately non-Hispanic White) and therefore under a specific set of assumptions, basing an allocation on race/ethnicity could lead to more deaths in each race/ethnicity stratum than basing it on age, even though race/ethnicity is a very strong predictor of mortality. A solution to avoid this issue is to maximize the outcome over the strongest predictor (age in this example), or ideally, to stratify by multiple predictors, as done by others \cite{chapman_risk_2022, wrigley-field_geographically_2021, kipnis_evaluation_2021, rosenstrom_can_2022, stafford_retrospective_2023}.

    \item The studies identified show diverse examples of how models can address equity considerations. "Equity" and related concepts mean different things to different researchers. For example, some publications emphasize equity as it relates specifically to the outcomes in the model and attempt to minimize inequality measures (which may be relative \cite{rosenstrom_can_2022} or absolute \cite{aiona_disparate_2022, kipnis_evaluation_2021}. Others note categories of preexisting disadvantage due to racism or geography, for example, and focus on how to improve outcomes for already-disadvantaged groups \cite{kadelka_ethnic_2022, chen_strategic_2022, stafford_retrospective_2023}. A way to improve the comparability of model predictions would be to present absolute outcomes (number of lives saved or lost, for example) and allocations (number of vaccines administered, for example) by subgroup, allowing a reader to calculate disparity measures of particular interest, which may be relative or absolute, and may focus on particular subgroups of the population. \\ 
    To facilitate interpretation by different readers according to multiple conceptions of equity, it may be advisable to present absolute outcomes (e.g., cases, deaths) overall and in each group, rather than or in addition to presenting results on the differences between groups. For example, Wrigley-Field et al. (2021) show on the same plots both the death rate in the population, as well as the proportion of BIPOC eligible for vaccination \cite{wrigley-field_geographically_2021}.

    \item Care should be taken in model input not to include measures that reflect already existing inequalities. For example, although life expectancy may vary by races/ethnicities \cite{arias_united_2022}, an average life expectancy should be used for all individuals of a given age in order to avoid downgrading the value of saving a life of a person of a certain age in already disadvantaged group \cite{rumpler_fairness_2023}.

    \item Most COVID-19 vaccine allocation studies were conducted and published in the face of significant uncertainty about pathogen characteristics and transmission dynamics. Most of the papers we found \cite{wrigley-field_geographically_2021, chapman_risk_2022, kadelka_ethnic_2022, kipnis_evaluation_2021, stafford_retrospective_2023, khodaee_humanitarian_2022} dealt with this by showing the robustness of their results across sensitivity analyses, thus strengthening the argument for the allocation policies they identified as best. For example, Kadelka et al. (2022), replicated all their results varying scenarios defined by three parameters of interest (ethnic homophily, relative contact level in high- vs low-contact jobs, and relative proportion of people of color compared to White people in high-contact jobs) \cite{kadelka_ethnic_2022}. 

    \item Some of the papers considered only the direct protection from vaccination on those receiving it, not the indirect effect on the rest of their communities \cite{rumpler_fairness_2023, wrigley-field_geographically_2021, chapman_risk_2022}. Doing so can be tempting, as indirect effects may not happen and can be more difficult to model with certainty, especially at the time allocation decisions are being made. Direct protection of vaccination is always a useful metric and may be more robust to assumptions on vaccine protection against infection or emergence of new variants. However, as highlighted by Jia et al. (2024), direct effects may not necessarily represent a lower bound of overall effects of vaccination, for example when vaccine wanes or when transmission or fatality parameters increase over time \cite{jia_causal_2024}.

\end{itemize}

\section{Acknowledgements}
We thank Carrie Wade from Countway Library, Harvard Medical School for her help in refining the Pubmed search. \\ 
\noindent We thank the four experts that answered our expert-opinion elicitation. \\

\newcolumntype{L}{>{\centering\arraybackslash}m{3cm}}

\newpage
\newgeometry{margin=1cm}
\begin{landscape}

\begin{table}[!ht]
    \centering
    \resizebox{26cm}{!}{\begin{tabular}{|V{2cm}|V{1.4cm}|V{3.5cm}|V{3cm}|V{5cm}|V{3.6cm}|V{3cm}|V{1.2cm}|V{2.3cm}|V{1cm}|}
    \hline
        \# & First author & Title & Study population  & What choice set is considered ?  & Consider heterogeneity in the population by which factor ?  & Which variable is being optimized ? & Explicitly mentions equity ?  & Consider equity by which factor ? & Link  \\ \hline
        1 & Aiona & The disparate impact of age-based COVID-19 vaccine prioritization by race/ethnicity in Denver, Colorado & Denver, Colorado & Does not model different interventions, shows the impact per race/ethnicity of age-based vaccination strategy over the three vaccination phases & Age and race/ethnicity & Nothing is optimized, they show the impact per race/ethnicity of age-based vaccination strategy & Yes & Race/ethnicity & https://pubmed.ncbi.nlm.nih.gov/35892113/ \\ \hline
        2 & Chapman & Risk factor targeting for vaccine prioritization during the COVID-19 pandemic & California & Once HCW and residents of LTCF are vaccinated, compares 8 strategies: i) random, ii) special pop, iii) age, iv) essential worker, v) comorbidity, vi) age and county, vii) age and special pop, viii) optimal using all factors & Age, geography (county), comorbidities, sex, special populations like prisoners, homeless, essential workers frontline or not and education & Cases, deaths, DALYs & No & Does not explicitely consider equity, but does present results by subgroups & https://pubmed.ncbi.nlm.nih.gov/35197495/ \\ \hline
        3 & Chen & Strategic COVID-19 vaccine distribution can simultaneously elevate social utility and equity & Nine (9) large metro areas (MSA) in the US (Atlanta Chicago Dallas Houston Los Angeles Miami Philadelphia San Francisco Washington, DC) & Compares 8 vaccine distriu=bution strategies: prioritize by age, prioritize by income, prioritize by occupation, prioritize by race/ethnicity, SVI-informed, comprehensive-ablation, homogeneous, comprehensive & Geography (9 cities), age, income, occupation, race/ethnicity, mobility data & Deaths & Yes & Age, income, occupation, race/ethnicity & https://pubmed.ncbi.nlm.nih.gov/36008683/ \\ \hline
        4 & Hoover & California’s COVID-19 Vaccine Equity Policy: Cases, Hospitalizations, And Deaths Averted In Affected Communities & California (analysis of the state's vaccine equity policy) & Shows the distribution of burden before and after vaccine equity policy is implemented, and counterfactual outcome had it not been implmented & Socioeconomic measure (Health Places Index), which represents ability to live healthy life & Vaccines doses and cases (and hospitalization and deaths in Supplementary) & Yes & Geography (ZIP code level Healthy Places Index) & https://pubmed.ncbi.nlm.nih.gov/38709962/ \\ \hline
        5 & Kadelka & Ethnic homophily affects vaccine prioritization strategies & US whole country & Shows optimal allocation for 6 different outcomes (deaths and cases overall and death and cases per racial/ethmicity groups NH White or BIPOC) & Age, race, and occupation & Cases, deaths, overall and for racial groups (WA or POC) & Yes & Race/ethnicity & https://pubmed.ncbi.nlm.nih.gov/36208667/ \\ \hline
        6 & Khodaee & A humanitarian cold supply chain distribution model with equity consideration: The case of COVID-19 vaccine distribution in the European Union & European Union coutries & Vaccine allocation to geographies (AR affected regions), here countries of the European Union & Characteristics of each geographies/countries (population, demand, etc) & Logistic costs (transportation, shortage, deprivation and witholding costs) and social costs (infections and deaths rates) & Yes & Geography ( affected regions AR) & https://www.sciencedirect.com/science/article/pii/S2772662222000571 \\ \hline
        7 & Kipnis & Evaluation of Vaccination Strategies to Compare Efficient and Equitable Vaccine Allocation by Race and Ethnicity Across Time & California (Kaiser Permanente Northern California patients) & Consider 6 scenarios: random, cdc proxy, age, risk, provid, crs & Age, sex, comorbidities, race & Cases, deaths, hospitalization, vaccine doses & Yes & Race/ethnicity, geography (high neigborhood deprivation index) & https://pubmed.ncbi.nlm.nih.gov/35977198/ \\ \hline
    \end{tabular}}

\end{table}

\end{landscape}
\restoregeometry

\newpage

\newgeometry{margin=1cm}
\begin{landscape}

\begin{table}[!ht]
    \centering
    %\resizebox{\textwidth}{!}
    \resizebox{26cm}{!}{\begin{tabular}{|V{2cm}|V{1.4cm}|V{3.5cm}|V{3cm}|V{5cm}|V{3.6cm}|V{3cm}|V{1.2cm}|V{2.3cm}|V{1cm}|}
    \hline
        \# & First author & Title & Study population  & What choice set is considered ?  & Consider heterogeneity in the population by which factor ?  & Which variable is being optimized ? & Explicitly mentions equity ?  & Consider equity by which factor ? & Link  \\ \hline
        8 & Munguía-López & Fair Allocation of Potential COVID-19 Vaccines Using an Optimization-Based Strategy & Mexico states & Optimization according to one of the four allocation schemes (Social welfare, Rawlsian justice, Nash scheme, Social welfare II) & Characteristics of Mexico tates (number beds, mortality, number high risk people, population etc) & Social welfare scheme maximizes total utility (sum of individual utilities), Rawlsian justice scheme maximizes the utility of the least well-off state, Nash scheme maximize product of total utilities & Yes & Geography (states) & https://www.ncbi.nlm.nih.gov/pmc/articles/PMC7804910/ \\ \hline
        9 & Rosenstrom & Can vaccine prioritization reduce disparities in COVID-19 burden for historically marginalized populations? & North Carolina & Compares: age then essential worker (EW), or EW then age, or no prioritization. Also tests increasing uptake in black/hispanic. Also looks at impact of susceptible only vaccination & Race/ethnicity. And the outcomes are all age-adjusted & Cases, deaths, hospitalization, as well as disparity (difference between death rate in Black/Hispanic/Other compared to White people)  & Yes & Race/ethnicity & https://www.ncbi.nlm.nih.gov/pmc/articles/PMC9801966/\#:\~:text=Using\%20a\%20simulation\%2Dbased\%20model,may\%20further\%20exacerbate\%20prevaccine\%20disparities. \\ \hline
        10 & Rumpler & Fairness and efficiency considerations in COVID-19 vaccine allocation strategies: A case study comparing front-line workers and 65–74 year olds in the United States. & US whole country and each state & After vaccinating HCW, nursing home residents, and 80+ yo, compares performance overall and per racial/ethnic group of vaccinating flw of all ages or 65-74 yo & Race/ethnicity, US state, age and front-line worker occupation & Deaths, YLL & Yes & Race/ethnicity, state & https://www.ncbi.nlm.nih.gov/pmc/articles/PMC10021220/ \\ \hline
        11 & Stafford & Retrospective analysis of equity-based optimization for COVID-19 vaccine allocation & Oregon & Compares any distribution of vaccines between two race groups (NH White and BIPOC) and four age groups  & Age and race (NH White or BIPOC) & Either burden (cases or deaths), or inequity (5 different definitions, described in Table 1), or both (5 combinaisons, as shown in Figure 3) & Yes & Race/ethnicity, age & https://academic.oup.com/pnasnexus/article/2/9/pgad283/7265347 \\ \hline
        12 & Wrigley-Field & Geographically targeted COVID-19 vaccination is more equitable and averts more deaths than age-based thresholds alone & California and Minnesota & Four sets of analyses: 1) age based alone 2) age based + deprivation geographies that have higher mortality rate than lower age band allowed 3) high mortality neighborhoods universal age 4) comparing lowering age threshold for high mortality geographies versus universally lowering age threshold & Age, race/ethnicity, geography (metro or non-metro and deprivation category, or historic covid mortality) & Deaths, YLL & Yes & Age, and race/ethnicity, geography & https://pubmed.ncbi.nlm.nih.gov/34586843/ \\ \hline
    \end{tabular}}
    \caption{Literature review}
    \label{tab:litreview}
\end{table}

\end{landscape}
\restoregeometry

\newpage

\newpage 
\printbibliography
\newpage

\end{document}